\documentclass[prl,nofootinbib
]{revtex4-2}

\usepackage{amsmath,amssymb,bm,graphicx}
\usepackage{multirow}

\usepackage{dcolumn}
\usepackage{latexsym}
\usepackage[dvipdfmx, colorlinks=true, linkcolor=blue]{hyperref} 
\def\<{\langle}
\def\>{\rangle}

\newcommand{\LA}{\mathop{\mathcal{L}}\nolimits}
\newcommand{\HA}{\mathop{\mathcal{H}}\nolimits}
\newcommand{\D}{\mathop{\mathcal{D}}\nolimits}

\newcommand{\I}{\mathop{\mathbb{I}}\nolimits}
\newcommand{\R}{\mathop{\mathbb{R}}\nolimits}
\newcommand{\CA}{\mathop{\mathbb{C}}\nolimits}
\newcommand{\ketbra}[2]{| #1 \rangle \langle #2 | }
\newcommand{\bracket}[1]{\langle #1  \rangle}

\newcommand{\bra}[1]{\langle #1 |}

\newcommand{\ket}[1]{| #1 \rangle} 
\renewcommand{\Re}{\mathop{\mathrm{Re}}}

\newtheorem{Thm}{Theorem}

\begin{document}
\title{Universal Constraints on Relaxation Times for $d$-level GKLS master equations}
\author{Gen Kimura, Shigeru Ajisaka, and Kyouhei Watanabe}
\affiliation{
Shibaura Institute of Technology, Saitama,  330-8570, Japan
}
\begin{abstract}In 1976, Gorini, Kossakowski, Sudarshan and Lindblad independently discovered a general form of master equations for the open quantum Markovian dynamics. 
In honor of all the authors, the equation is nowadays called the GKLS master equation.  
In this paper, we show universal constraints on the relaxation times valid for any $d$-level GKLS mater equations, which is a generalization of the well-known constraints for $2$-level systems. 
Specifically, we show that any relaxation rate, the inverse-relaxation time, is not greater than the half of the sum of all relaxation rates. 
Since the relaxation times are measurable in experiments, our constraints provide a direct experimental test for the validity of the GKLS master equations, and hence for the conditions of the completely positivity and Markovianity. 
\end{abstract}


\maketitle


\section{Introduction}

The study of the open quantum dynamics holds significant value both in fundamental theory and technologies \cite{ref:AL,ref:P,ref:GZ,ref:GKreview}. 
On the one hand, the origin of the non-unitary dynamics has been in debate for many
years, especially in the connection with the measurement problem \cite{ref:des}. 
On the other hand, the decoherence is the key obstacle in achieving practical quantum devices such as the quantum computer \cite{ref:NC}. 
Therefore, a better understanding of the nature of decoherences and relaxation processes in open quantum systems is becoming almost an urgent task for theorists due to demands from quantum technologies. 

While a general property of a time evolution is the unitarity (reversibility) for isolated quantum systems, 
it is widely accepted that the {\it completely positivity} plays such a role for an open quantum system \cite{ref:Kraus}. 
The completely positivity comes from the the positivity preservation for a system coupled
to an environment owning finite dimension. Indeed, one can directly prove the completely positive condition of a time evolution map under any interaction between the system $S$ and the environment $E$, provided that there is no initial correlations between $S$ and $E$ \cite{ref:GKS}. 
However, some of the authors believe that further debate is necessary before setting the completely positivity as an axiom of the theory of open quantum dynamics, especially in connection with the presence of initial correlation \footnote{
We do not go any further into this problem here. However, it deserves to be noticed that the real problem lies not in the completely positive condition but in the validity of the concept of time evolution map (See the footnotes 29 and 30 in \cite{ref:GKreview}). }. 
Therefore, it would be quite prominent if there is a direct way to check the validity of the completely positivity in experiments. 

In many physical situations, especially where the interaction with an environment is not too strong, the dynamics is well described by the Markovian dynamics where the memory effect turns out to be negligible. (Note that the validity of the assumption can be quantitatively studied by measuring the correlation times in environments.) \cite{ref:P,ref:GZ}.  
The time evolution map $\Lambda_t$ is then described by a one parameter $t \in \R$ semigroup acting on the set of density operators:
$$
\rho \to \rho_t = \Lambda_t \rho, 
$$
which is called a (completely positive) dynamical semigroup \cite{ref:DS,ref:GKS}. 
With a natural continuity condition, we get the master equation for the density operator
\begin{equation}\label{eq:meq}
\frac{d}{dt} \rho_t = \LA \rho_t,
\end{equation}
where $\LA$ is called the generator of the dynamics \cite{ref:Yoshida}.   
In 1976, Gorini, Kossakowski, and Sudarshan \cite{ref:GKS}, and independently Lindblad \cite{ref:Lind} were successful in obtaining a general form of the generator $\LA$ for any completely positive dynamical semigroups: If the system of interest is a quantum system with the associated $d$-dimensional Hilbert space, the generator $\LA$ can be decomposed into the Hamiltonian part $\HA$ and the dissipative part $\D$ as 
\begin{eqnarray}\label{eq:GKLS}
\LA = \HA + \D, \ \HA (\rho) =-i[H,\rho], \ \D(\rho) = \frac{1}{2} \sum_{i,j=1}^{d^2-1} C_{ij} ([F_i,\rho F_j^\dagger] + [F_i\rho, F_j^\dagger]) 
\end{eqnarray}
where $H$ is an Hermitian matrix (an effective Hamiltonian), $[C_{ij}]$ is a positive matrix, and $(F_i)_{i=1}^{d^2}$ with $F_{d^2} = \I/\sqrt{d}$ is an orthonormal basis of the set of all linear operators with respect to the Hilbert-Schmidt inner product, and $[A,B]:= AB -BA$ denotes the commutator between linear operators $A,B$ \footnote{The generator can be written for the infinite dimensional system as well \cite{ref:Lind}. However, in the present paper, we restrict ourselves to finite dimensional cases. }. 
Equation (\ref{eq:meq}) with the form (\ref{eq:GKLS}) is called the GKLS (Gorini-Kossakowski-Lindblad-Sudarshan) master equation in honor of all the authors for this seminal discovery, and is widely used in the variety of fields such as quantum optics, quantum information sciences, biology, condensed matter physics and particle physics \cite{ref:AL,ref:P,ref:GZ}.  
As essential assumptions for the GKLS master equations are the completely positivity and Markovianity, it is quite interesting to ask a general physical character of GKLS master equations directly available in experiments.     

The authors in \cite{ref:GKS} indeed discussed such a general feature and revealed a generic constraints among the relaxation times in two level system, i.e., a qubit system: The three relaxation times $T_i \ (i = 1, 2, 3)$ of the two level systems are shown to satisfy the following inequalities
\begin{equation}\label{eq:2costraint}
1/T_i + 1/T_j \ge 1/T_k, \ (i,j,k): \ {\rm permutation \ of} \ (1,2,3)
\end{equation} 
In particular, these relations includes the famous relation between the longitudinal relaxation time $T_L (=T_1)$ and the transverse relaxation time $T_T (=T_2 = T_3)$:
\begin{equation}
2 T_L \ge T_T.   
\end{equation} 
This relation is usually observed in experiments \cite{ref:AL,ref:LT}, the fact of which reflects the universal validity of GKLS master equations in open quantum systems.  
While the authors in \cite{ref:GKS} discussed the relations (\ref{eq:2costraint}) only for a restricted setting where the Hamiltonian part $\HA$ and $\D$ are commutative, which is equivalent to the case of Pauli master equations \cite{ref:GKDoc}, the constraints (\ref{eq:2costraint}) was later shown to hold for arbitrary GKLS master equation in two level system \cite{ref:GK}. 
For higher level systems, Schimer and Solomon has studied some properties for relaxation times \cite{ref:SS}, but to the best of the authors' knowledge, there are no universal constraints which are known to hold for any GKLS master equations. 
The purpose of this paper is to generalize constraints (\ref{eq:2costraint}) to an arbitrary $d$-level quantum system. 
Namely, for generally $d^2-1$ existing relaxation times $T_\alpha \ (\alpha = 1,\ldots,d^2-1)$ (see next section for the detail) of $d$-level GKLS master equations, it follows that 
\begin{equation}\label{eq:Gconst}
\frac{1}{2} \sum_{\beta=1}^{d^2-1} \frac{1}{T_\beta} \ge \frac{1}{T_\alpha}. 
\end{equation}
One observes easily that this is a generalization of (\ref{eq:2costraint}) which is the case of $d=2$. 
Naively speaking, the constraints (\ref{eq:Gconst}) states that GKLS master equation, or the completely positive condition, prohibits the case where only one of the decaying time scale is too small. 
Indeed, we shall prove more tight constraints than (\ref{eq:Gconst}) for $d \ge 3$ in the next section (See Theorem \ref{thm:Main}). 

\section{Universal Constraints on Relaxation Times for $d$-level systems}

In the following, let the system of interest be an arbitrary $d$-level quantum system associated with the Hilbert space $\CA^d$. 
The Hilbert-Schmidt inner product between two matrices $A,B \in M_d(\CA)$ is denoted by $\bracket{A,B} := \mathrm{Tr} (A^\dagger B)$. 

Let us start from clarifying the concept of the relaxation time in $d$ level GKLS master equations. 
Mathematically, GKLS master equation is an ordinary differential equation of the first order and dimension $d^2$, and the general solutions are superposition of the terms $P(t) \exp(\lambda t)$ where $P(t)$ is a polynomial function of $t$ and $\lambda$ is an eigenvalue of the generator $\LA$. 
Considering the trace preserving property, one of the eigenvalue of $\LA$ is $0$\footnote{
Let $\Lambda^\ast$ be the adjoint of $\Lambda$, i.e, $\bracket{\LA^\ast(A),B} = \bracket{A,\LA(B)}$ for $A,B \in M_d(\CA)$. 
From the trace preserving property, one has $0 = \frac{d}{dt} \mathrm{Tr} (\rho_t)|_{t=0} = \mathrm{Tr} \LA(\rho) = \mathrm{Tr} \LA^\ast(\I) \rho$ for any $\rho := \rho_0 \in M_d(\CA)$, so it follows that $\LA^\ast(\I) = 0$. As the adjoint map has conjugate eigenvalues in general, so $\LA$ has also a zero eigenvalue. }, so there are in general $d^2-1$ other eigenvalues $\lambda_\alpha \ (\alpha=1,\ldots,d^2-1)$. 
Moreover, the real part of any eigenvalues of $\LA$ cannot be positive, since otherwise the corresponding solution eventually go outside the set of all density operators, which is the bounded set, with e.g., the Hilbert-Schmidt norm. 
With these facts in mind, we define $d^2-1$ relaxation rates by $\Gamma_\alpha: = - \Re \lambda_\alpha > 0 \ (\alpha=1,\ldots,d^2-1)$, and the relaxation times by $T_\alpha := 1/\Gamma_\alpha$. 
By the linearity of $\LA$ and the Born's rule of quantum mechanics, an expectation value (including a probability) of any physical quantity $A$ evolves as the superposition of the exponential decay with these time scales:
$$
\bracket{A}_t = \sum_{\alpha = 1,\ldots,d^2-1} P_\alpha(t) \exp(- t/T_\alpha) \exp(i \omega_\alpha t) + C,   
$$
where $P_\alpha(t)$ is a polynomial of $t$, $C$ is a constant (corresponding to the $0$ eigenvalue) and $\omega_\alpha := \Im \lambda_\alpha$ for the oscillating terms. 
Therefore, by measuring various quantities $A$, one can in principle observe the relaxation times $T_\alpha$ as exponential decaying time scales. 

We are now ready to prove the main theorem:
\begin{Thm}\label{thm:Main}
For any GKLS master equation in $d$-level quantum system, all relaxation rates $\Gamma_\alpha \ (\alpha = 1,\ldots,d^2-1)$ satisfy 
\begin{equation}\label{eq:main}
\sum_{\beta=1}^{d^2-1} \Gamma_\beta \ge \frac{d}{\sqrt{2}} \Gamma_\alpha. 
\end{equation}
\end{Thm}
Note that $\frac{d}{\sqrt{2}}$ for all $d\ge 3$, constraints (\ref{eq:main}) implies (\ref{eq:Gconst}) by using $T_\alpha = 1/\Gamma_\alpha$. As the case for $d=2$ has already shown to be true in \cite{ref:GK}, we conclude that the constraints (\ref{eq:Gconst}) universally follows for any $d$-level GKLS master equations.

[Proof] 
First, note that all complex eigenvalues of $\LA$ appears as conjugate pairs. 
To see this, note that the Positive and hence Hermiticity Preserving property of dynamical semigroup implies 
\begin{equation}\label{eq:HP}
(\LA(A))^\dagger = \LA(A^\dagger).
\end{equation} 
Let $u_\alpha \ (\alpha=1,\ldots,d^2-1)$ be the normalized eigenvector of $\LA$ belonging to the eigenvalue $\lambda_\alpha$, i.e, 
$\LA(u_\alpha) = \lambda_\alpha u_\alpha \ (||u_\alpha||^2 = \mathrm{Tr} u^\dagger_\alpha u_\alpha = 1).$
By taking the adjoint of this and using (\ref{eq:HP}), we have $\LA(u^\dagger_\alpha) = \overline{\lambda_\alpha} u^\dagger_\alpha$.  
This shows that if $\lambda$ is an eigenvalue of $\LA$, then its complex conjugate $\overline{\lambda}$ is also an eigenvalue of $\LA$. Combining this and the fact that one of eigenvalue of $\LA$ is $0$, we have 
\begin{equation}
\mathrm{Tr} \LA = - \sum_{\beta=1}^{d^2-1} \Gamma_\beta. 
\end{equation}
Moreover, the trace of GKLS generator $\LA$ can be directly computed to be $-d \mathrm{Tr} C$ \footnote{
Using ONB $(\ketbra{k}{l})_{k,l=1}^d$ of $M_d(\CA)$, one has  
$$
\mathrm{Tr} \HA = \sum_{k,l} \bracket{\ketbra{k}{l}, \HA(\ketbra{k}{l})} = -i \sum_{k,l} (\bra{k}H\ket{k} - \bra{l}H\ket{l}) = 0, 
$$
and 
$$
\mathrm{Tr} \D = \frac{1}{2} \sum_{i,j} C_{ij} \sum_{k,l} (2 \bra{k}F_i \ket{k}\bra{l}F_i \ket{l} - \bra{k} F_j^\dagger F_i \ket{k} - \bra{l} F_j^\dagger F_i \ket{l})$$
$$
= \frac{1}{2} \sum_{i,j} C_{ij} (2 \mathrm{Tr}  F_i \mathrm{Tr} F_j - 2 d \mathrm{Tr} F_j^\dagger F_i ) = - d \sum_{i} C_{ii} = -d \mathrm{Tr} C, 
$$
where we have used that $\mathrm{Tr} F_i = \bracket{F_{d^2},F_i} = 0$ and $\mathrm{Tr} F_i^\dagger F_j = \bracket{F_i,F_j} = \delta_{ij}$. 
}, 
we have  
\begin{equation}\label{eq:trCG}
\sum_{\beta=1}^{d^2-1} \Gamma_\beta = d \mathrm{Tr} C,
\end{equation}
where $C$ is the positive matrix in (\ref{eq:GKLS}). 

By GKLS form (\ref{eq:GKLS}), the eigenvalue equation for $\LA$ reads 
\begin{equation}\label{eq:Lu}
-i [H, u_\alpha] + \frac{1}{2} \sum_{ij} C_{ij} (2 F_i u_\alpha F_j^\dagger - \{F_j^\dagger F_i, u_\alpha\} ) = \lambda_\alpha u_\alpha 
\end{equation}
and its conjugation 
\begin{equation}\label{eq:Lud}
-i [H,u^\dagger_\alpha] + \frac{1}{2} \sum_{ij} C_{ij} (2 F_i u^\dagger_\alpha F^\dagger_j - \{F_j^\dagger F_i, u^\dagger_\alpha\} ) = \overline{\lambda_\alpha} u^\dagger_\alpha. 
\end{equation}
Here $\{A,B\}:= AB + BA$ denotes the anti-commutator between linear operators $A,B$. 
Multiplying $u_\alpha^\dagger$ to (\ref{eq:Lu}) from the left and $u_\alpha$ to (\ref{eq:Lud}) from the right, we have 
\begin{eqnarray}
-i u^\dagger_\alpha[H, u_\alpha] + \frac{1}{2} \sum_{ij} C_{ij} (2 u^\dagger_\alpha F_i u_\alpha F_j^\dagger - u^\dagger_\alpha\{F_j^\dagger F_i, u_\alpha\} ) = \lambda_\alpha u^\dagger_\alpha u_\alpha \\
-i [H,u^\dagger_\alpha]u_\alpha + \frac{1}{2} \sum_{ij} C_{ij} (2 F_i u^\dagger_\alpha F^\dagger_j u_\alpha- \{F_j^\dagger F_i, u^\dagger_\alpha\}u_\alpha ) = \overline{\lambda_\alpha} u^\dagger_\alpha u_\alpha
\end{eqnarray} 
Taking traces over both equations and summing them up, we have 
$$
\sum_{ij} C_{ij} \mathrm{Tr} (u^\dagger_\alpha F_i u_\alpha F_j^\dagger + F_i u^\dagger_\alpha F^\dagger_j u_\alpha - u^\dagger_\alpha u_\alpha F_j^\dagger F_i -  u_\alpha u^\dagger_\alpha F_j^\dagger F_i) = -2 \Gamma_\alpha,  
$$ 
where we have used $2 \Re \lambda_\alpha = - \Gamma_\alpha$ and $\mathrm{Tr} u^\dagger_\alpha u_\alpha =1$ (the normalization condition for eigenvectors). 
Using the eigenvalue decomposition of the positive matrix $C$, $C_{ij} = \sum_k p_k v^{(k)}_i \overline{v^{(k)}_j} $ with $p_k \ge 0$ and letting $L_k:= \sum_i v^{(k)}_i F_i$, we have 
\begin{equation}
\sum_{k} p_k \mathrm{Tr} (u^\dagger_\alpha L_k u_\alpha L_k^\dagger + L_k u^\dagger_\alpha L^\dagger_k u_\alpha - u^\dagger_\alpha u_\alpha L_k^\dagger L_k -  u_\alpha u^\dagger_\alpha L_k^\dagger L_k) = -2 \Gamma_\alpha. 
\end{equation}
This can be rewritten as 
\begin{equation}
\sum_k p_k (\bra{[L_k,u_\alpha]}\ket{L_k u_\alpha} + \bra{[L_k,u^\dagger_\alpha]}\ket{L_k u_\alpha^\dagger}) = 2 \Gamma_\alpha. 
\end{equation}
By Schwarz inequality and the triangle inequality, 
\begin{equation}
2 \Gamma_\alpha \le \sum_k p_k (||[L_k,u_\alpha]|| ||L_k u_\alpha|| + || [L_k,u^\dagger_\alpha] || || L_k u_\alpha^\dagger||) \le \sum_k p_k (||[L_k,u_\alpha]|| + || [L_k,u^\dagger_\alpha] || ) ||L_k||,
\end{equation}
where we have used $||A B|| \le ||A|| ||B||$ and $||u_\alpha|| = ||u^\dagger_\alpha|| = 1$. 

Finally, using the inequality \cite{ref:BW}
$$
|| [A,B]|| \le \sqrt{2} ||A|| ||B||, 
$$ 
we obtain 
\begin{equation}
\Gamma_\alpha \le \sqrt{2}\sum_k p_k ||L_k|| = \sqrt{2} \sum_k p_k =  \sqrt{2} \mathrm{Tr} C, 
\end{equation} 
where we have used $||L_k||^2 = \mathrm{Tr} L_k^\dagger L_k = \sum_{ij} (\sum_i \overline{v^{(k)}_i}v^{(k)}_j F_i ^\dagger F_j = \sum_{i} |v^{(k)}_i|^2 = ||v^{(k)}||^2 = 1$. 
Hence, 
\begin{equation}
\frac{d}{\sqrt{2}} \Gamma_\alpha \le \sum_{\alpha = 1}^{d^2-1} \Gamma_\alpha. 
\end{equation}
Applying (\ref{eq:trCG}), we obtain the constraints (\ref{eq:main}). 

\hfill $\blacksquare$ 

\section{Conclusion and discussion}
In this paper, we have investigated $d$-level GKLS master equations and obtained universal constraints (\ref{eq:Gconst}) for relaxation times that are predicted by any GKLS master equation. 
Indeed, the obtained constraints are more tight for $d \ge 3$ and are expressed as (\ref{eq:main}).
As the relaxation times are in principle measurable in experiments, the constraints would serve as a direct check for the validity of GKLS master equation, or equivalently, for the validity of completely positive conditions and Markovianity.

\section*{Acknowledgments}
We would like to thank Prof. D. Chru\'sci\'nski and Prof. K. \.{Z}yczkowski for their useful comments and discussions. 
G.K. is grateful to Prof. Kossakowski for inviting me to this interesting subject and for supporting my work with helpful and fruitful advices since he was a postdoc at his laboratory.

\end{document}